# Optical study of stress hormone-induced nanoscale structural alteration in brain using partial wave spectroscopic (PWS) microscopy

Shiva Bhandari,[1] Pradeep Shukla,[2] Huda Almabadi,[1] Peeyush Sahay,[1] Radhakrishna Rao,[2] Prabhakar Pradhan[1*]

[1]Department of Physics and Materials Science, BioNanoPhotonics Laboratory, University of Memphis, Memphis, Tennessee, 38152, USA
[2]Department of Physiology, University of Tennessee Health Science Center, Memphis, TN 38103, USA

*P. Pradhan : ppradhan@memphis.edu

## Abstract

Chronic stress affects nano to microscale structures of the brain cells/tissues due the suppression of neural growths and reconnections, hence the neuronal activities. This results in depression, memory loss and even the death of the brain cells. Our recently developed novel optical technique, partial wave spectroscopic (PWS) microscopy has nanoscale sensitivity, and hence, can detect nanoscale changes in brain tissues due to stress. In this study, we applied this technique to quantify the stress related structural changes in the corticosterone-treated mouse model of stress. Our results show that brains from corticosterone-treated mice showed higher nanoscale structural disorder in the hippocampal region as compared to the brain from normal (vehicle) mice. The increase in structural alteration correlates with the duration of the stress. We further quantified the relative changes and the spatial localization of these changes in this mouse model and found out that the maximum changes occurred nearly symmetrically in both regions of the hippocampus. The mRNA for stress-related genes, BDNF and TrkB were also significantly reduced in the hippocampus of corticosterone-treated mice compared to that in control mice. These results indicate that chronic corticosterone treatment induces nanoscale structural alterations in mouse brain that corresponds to the changes in stress-related gene expression.

# 1. Introduction:

Stress has become an inevitable part of modern life. The hormone cortisol is mainly responsible for stress in humans. There are mainly two types of stresses: acute and chronic. When we talk about stress, it does not always have negative effects on brain. Some stress is necessary for the active and cognitive performance and development. Research have shown that acute stress helps in the proliferation of neural stem cells into new nerve cells and these new nerve cells on maturation increases the mental performance. The release of protein named fibroblast growth factor 2 (FGF2) by astrocytes in brain from stress causes the proliferation of these new nerve cells. Cortisol released because of acute stress is a response of fight-or-flight mechanism and the level of cortisol returns to original without any long-lasting effects. On other hand, chronic stress increases the level of stress hormone cortisol and this suppresses the development of new neurons in hippocampus as well as shrinkage of neural network. Chronic stress can lead to different body problems ranging from simple depression, viral infection to cancer. Most of the health problems nowadays are somehow inter-connected to stress [1,2,3]. Studies have shown that chronic stress is responsible for mental and physical disorder by altering the gene control [4].

Epinephrine, norepinephrine and cortisol are the main three stress hormones in humans. Epinephrine and norepinephrine are the flight and fight hormones mainly responsible for the immediate reactions we feel when stressed. These hormones make us more aware, awake and focused and help in fleeing away from the stressful condition. Cortisol is the stress hormone which helps in fluid balance and blood pressure in optimal level. However, excessive level of cortisol interferes with learning and memory, lowers the immune system, increase body weight, increase blood pressure, causes fluid imbalances, causes heart diseases, diabetes, etc. This stress hormone is also recognized as 'Public enemy #1'. [5, 6, 7]. Chronic stress has multiple effects on brain from forgetfulness to death of brain cells and many neurological disorders. The inter connection between the structural alteration of the brain cells/tissues and the effect of the stress hormone cortisol is not well known and requires a systematic study.

The role of brain-derived neurotrophic factor (BDNF) is established in stress and affective disorders. It has also been showed that BDNF was protective to neurons in conditions of chemical stress [8]. Both acute and chronic stresses have been shown to decrease the BDNF in various animal models and are involved in developing depressive phenotypes. Tyrosine kinase-coupled receptor (TrkB) is the primary signal transduction receptor for BDNF. BDNF and TrkB expression have been shown to decrease in depression patient hippocampus. The activation of the BDNF-TrkB pathway is important in the development and the growth of neurons.

In this study, we measured nanoscale structural alterations in brain using a mouse model of chronic corticosterone administration and the novel nanoscale optical imaging technique and gene expression analysis.

# 2. Method:

## 2.1. Corticosterone induced mice animal model and brain tissue collections:

The use of mice for experimentation purpose was approved by the Institutional Animal Care and Use Committee (IACUC) of the University of Tennessee UT, Health Science Center, Memphis, TN, USA, in accordance with institutional and U.S. federal guidelines. Mice were randomly divided into two groups (n = 6 per group) and injected (s.c.) either vehicle (control) or corticosterone (25 mg/kg in sesame oil) daily at 9 -10 AM for 7 and 10 days. Animals had ad libitum access to diet and water. Animals were sacrificed two hours after the injection on seventh and tenth day and brain hippocampus were collected for mRNA expression analysis. The brains were sectioned using a microtome and attached to a slide and fixed by using formalin. The thickness of each brain slice sample was 10 μm. The fixed slides were then used for partial wave spectroscopy (PWS) microscopy studies.

## 2.2. Optical partial wave experiments:

The instrumentation of the PWS system, developed at University of Memphis that provides finer focus, is shown in the Fig. 1. The sample is illuminated by a broadband of white light obtained from Xenon Lamp (Thor Lab) using Kohler Illumination. Light is first passed into a broadband dielectric round mirror (diameter= 50mm, R>99%, thickness=12.7mm,Newport) and is projected towards a set of lenses (convex, f=50mm,Thorlabs) and apertures for collimation as shown in Fig 1.A  right angle prism (25mm,Newport) reflects the collimated light and allows it  to pass through a beam splitter to a low numerical aperture objective (NA=0.65, 40x ,Newport). The beam is then focused on the sample. The sample is kept on the electronic motorized stage (Zaber Tech, Canada). The stage provides precise resolutions of 100nm in z direction and 40nm in x-y plane so that the control on the accurate depth of focus after reflection from the sample is obtained. The backscattered light from the sample is passed through the objective again and strikes into a liquid crystal tunable filter (LCTF) (Verispace, LLC). LCTF filters the signal according to the wavelength components at a resolution of 1nm within the visible range (450-700nm). A CCD camera detector coupled with the LCTF captures the filtered signal.

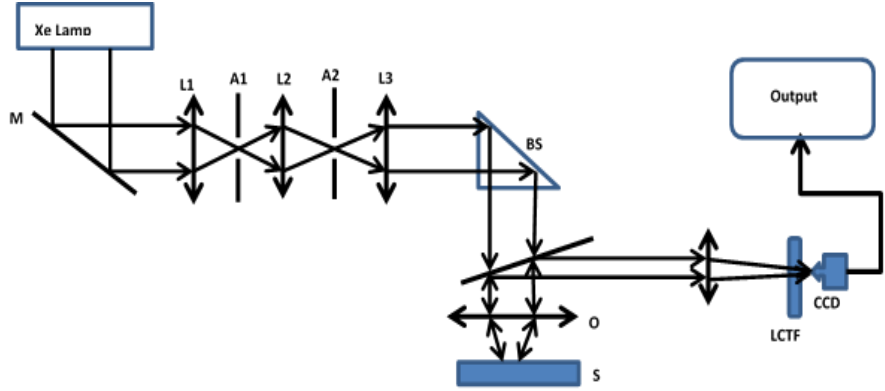

**Figure 1:** Experimental setup for the Partial Wave Spectroscopy system. L1, L2, L3 lens; A1, A2: Apertures; BS: Beam Splitter; M: Mirror; O Objective; S: Sample; LCTF: Liquid crystal Tunable Filter; CCD: Charge coupled device camera.

The backscattered image of the sample is detected and stored by the CCD camera at every wavelength ($\lambda$) in the visible range (450-700nm).. The backscattered light spectra at a visible wavelength range from 450 to 700nm are captured from each pixel of CCD. The backscattered light spectrum is then normalized by the spectrum of the incident light and filtered to remove spectral noise. This yields a data cube $R(\lambda, x, y)$ ($\lambda$ is the wavelength, x and y are the pixel coordinates), which is referred to as the fluctuating part of the reflection coefficient. PWS measures the spectral fluctuations in the backscattering spectra. PWS virtually divides a cell into a collection of back-scattered parallel channels, where each channel with a diffraction-limited transverse size, detects back-scattered waves propagating along quasi 1D trajectories within these channels, and quantifies the statistical properties of the nanoarchitecture of a cell by the analysis of the fluctuating part of the reflected intensity $R(\lambda,x,y)$. The reflection intensity spectral fluctuations in $R(\lambda,x,y)$ arise from the multiple interference of photons reflected from refractive index fluctuations within a scattering object, i.e., the cell. These spectral fluctuations, originated from the multiple scattering from the sample, are analyzed by means of a quasi 1D mesoscopic light transport theory. This enables the quantification of the statistical properties of the spatial refractive index variations at any length scale including those well below the diffraction limit. The statistical parameter obtained from the analysis is the disorder strength $L_d = <$ $\Delta n^2 > l_c$, where $<\Delta n^2>$ and $l_c$ are the variance and the spatial correlation length of the refractive index fluctuations [9]. The disorder strength quantifies the spatial variability of refractive index and thus, the local concentration of intracellular material.

Experiments were performed longitudinally on the slide containing the brain sections of the mice. Thirteen sets of PWS images for each three different samples were taken for each brain slices on the slide. The reflected back scattering spectra $R(\lambda,x,y)$ were taken for the whole visible range (400-700nm), and were analyzed using the mesoscopic scattering algorithm and the degree of the structural disorder, or disorder strengths were calculated [10]. The samples were mice brain tissue with vehicle (i.e., control), seven-day old mice brain tissue treated with corticosterone and ten-day old mice brain tissue treated with corticosterone. The average of the disorder strengths for each brain sample and the standard error were calculated the following formula:

$$L_d = <\Delta n^2> l_c$$

**Quantitative RT-PCR**

Total RNA (1.5 µg) was used for generation of cDNAs using the ThermoScript RT-PCR system for first strand synthesis (Invitrogen). Quantitative PCR (qPCR) reactions were performed using cDNA mix (cDNA corresponding to 35 ng RNA) with 300 nmoles of primers in a final volume of 25 µl of 2x concentrated RT2 Real-Time SYBR Green/ROX master mix (Qiagen) in an Applied Biosystems 7300 Real-Time PCR instrument (Norwalk, CT, USA). The cycle parameters were: 50°C for 2 minutes, one denaturation step at 95°C for 10 minutes and 40 cycles of denaturation at 95°C for 10 seconds followed by annealing and elongation at 60°C. Relative gene expression of each transcript was normalized to GAPDH using the ΔΔCt method. Sequences of primers used for qPCR are provided in Table 1.

**Table 1.**

| Gene | 5′-3′ Sequence |
|---|---|
| **TrkB** | |
| Forward Primer | CTGGGGCTTATGCCTGCTG |
| Reverse Primer | AGGCTCAGTACACCAAATCCTA |
| **BDNF** | |
| Forward Primer | TCATACTTCGGTTGCATGAAGG |
| Reverse Primer | AGACCTCTCGAACCTGCCC |
| **GAPDH** | |
| Forward Primer | CTGCACCACCAACTGCTTAG |
| Reverse Primer | GGGCCATCCACAGTCTTCT |

## 3. Results:

### 3.1. Partial wave optical experimental result of structural disorder in the stressed induced Brain tissue sections:

In Fig. 2, the PWS analysis of the different mice brain tissues slices are shown. It can be seen in the bar graphs that there is an increase in the nanoscale structural disorder strength parameter $L_d$ in the corticosterone treated mice brain tissues relative to the mice brain tissues with vehicle (control). The increase correlates with the duration of corticosterone treatments.

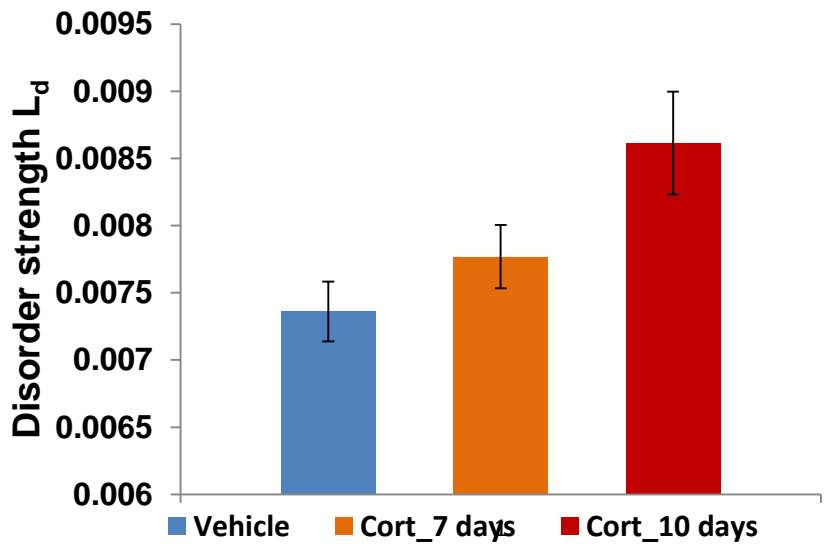

**Figure 2:** Figure: Stress-induced structural changes in mice brain tissues due to chronic stress treated with vehicle, brain tissue treated with corticosterone for seven days and brain tissue treated with corticosterone for ten days. The change in the disorder strength was found to be nearly 6% for brain tissues treated with corticosterone for seven days and nearly 17% for brain tissues treated with corticosterone for ten days. (P-value <0.05)

Corticosterone, a main glucocorticoid is responsible for stress in rodents [11]. PWS optical measurements show that the brain tissue structural disorder is increased in the corticosterone treated brain tissue relative to the control (vehicle). This confirms that there are structural alterations in the brain tissues after the injection of corticosterone. Stress is responsible for the nanoscale changes in the brain tissues slice samples. The percentage change in disorder relative to the control (vehicle) mice brain tissue was approx. 6% for seven-day corticosterone-treated mice brain tissue and approx. 17% for ten-day corticosterone-treated mice brain tissue. These results suggest that higher disorder strength in brain tissue is associated with the stress and hence the higher nanostructural alterations. The nanostructural alterations were found to be profound in the hippocampal region. Future study is needed to explore whether the disorder strength continues to increase with stress or saturates at a certain level of stress. In the following we studied the locations in brain where the higher nano structural alterations are taking place.

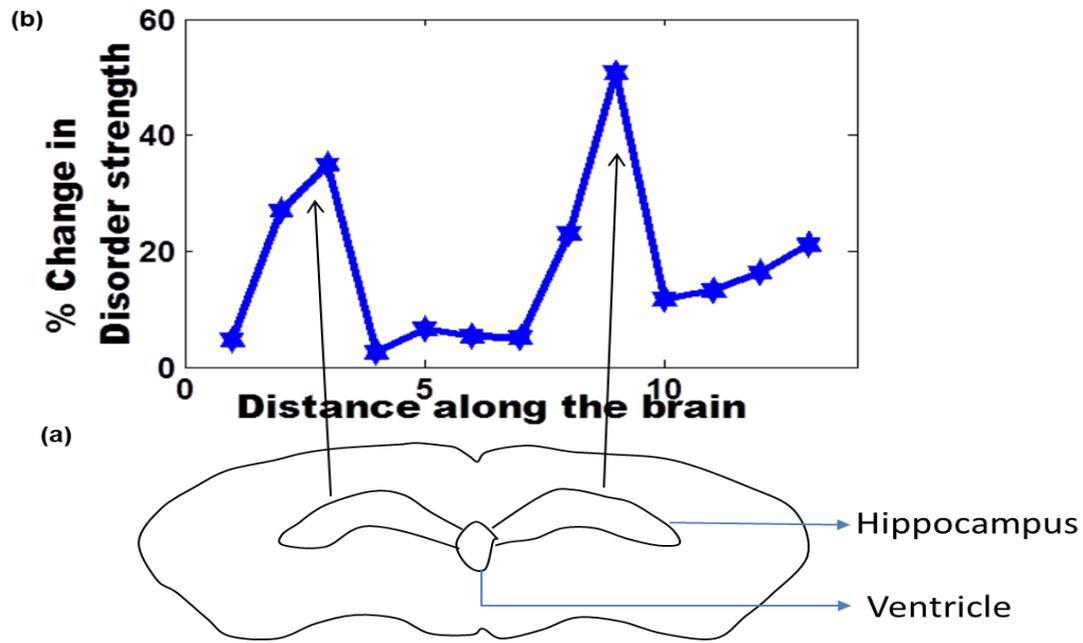

**Figure 3:** Change in structural disorder or disorder strength due to corticosterone induced chronic stress in mice brain. (a) Schematic outline of the mice brain sections with hippocampal region. The shaded area represents the area where the PWS experiment was performed. (b) Plot of the percentage change in the structural disorder or disorder strength $L_d$ versus position along the brain section (as shown in the shaded region). The two peaks in the graph represent the maximum changes occurring in the hippocampal regions which are symmetrical on either side of the ventricle.

## 3.2. The region specific structural changes in the brain:

PWS experiment was performed over 10µm thick brain section slide. The percentage change in the disorder strength along the brain slice sample position are maker by levels as shown in Fig. 3.(a). The middle portion in the figure is the ventricle and the region on either side of the ventricle is hippocampus. We plotted nano-structural changes in percentages along this shaded stripe shown in the figure. It can be seen in the figure that the percentage change versus position graph Fig. 3(b) clearly indicates that changes in disorder strength $L_d$ are more in the hippocampal region because of corticosterone induced stress.

### 3.3. Effect of corticosterone on BDNF and TrkB expression.

BDNF and TrkB are well known to be stress-related genes. We evaluated the expression of these genes in hippocampus by measuring mRNA levels. Data show that mRNA for both BDNF (Fig. 4(a) and TrkB (Fig. 4(b)) were significantly decreased in the hippocampus of corticosterone-treated mice compared to that in vehicle-treated mice at Day 7 and Day 10 in a duration-dependent manner.

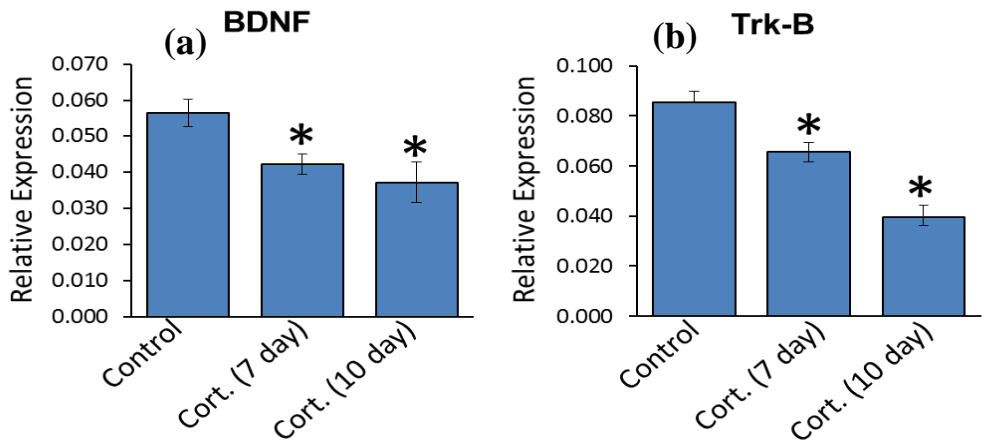

Figure 4: Corticosterone treatment down regulate BDNF and TrkB gene expression in the hippocampus. Stress was induced with corticosterone (25 mg/kg in sesame oil) daily for 7 and 10 days as described in methods. At the Day 7 and 10, qRT-PCR analysis performed in RNA from hippocampal tissue for BDNF (a) and TrkB (b). Values are mean ± SE ($n = 4$). *Asterisks* indicate the values that are significantly ($p < 0.05$) different from corresponding control values.

### 5. Discussions and Conclusions

In this paper, we have reported stress induced nanoscale structural changes in brain tissue of a mice model. The PWS optical study of the brain indicated that corticosterone-treated mice brain tissues have more structural changes than the mice brain tissues with vehicle. Corticosterone-treated day-ten mice brain tissues have more nanoscale structural disorder strength than corticosterone-treated day-seven brain tissues. The disorder strength increases with the increase in the duration of the corticosterone treatment. The changes are profound in the hippocampal region. The reversibility or irreversibility of the changes with the increase in corticosterone and total duration of injection is yet to be systematically explored.

The role of brain-derived neurotrophic factor (BDNF) has been implicated in stress and affective disorders. It has also been shown that BDNF is protective to neurons in conditions of chemical stress [12]. Both acute and chronic stresses have been shown to decrease BDNF expression in various animal models and are involved in developing depressive phenotypes. Our present study shows reduced BDNF expression in mouse hippocampus by chronic corticosterone administration and a corresponding increase in nanoscale structural disorder. Tyrosine kinase-coupled receptor (TrkB) is the primary signal transduction receptor for BDNF. BDNF and TrkB expression have been shown to decrease in the hippocampus of depression patients. The activation of the BDNF-TrkB pathway is important in the development and the growth of neurons. Suppression of this pathway by corticosterone suggests a chronically elevated corticosterone in mediating the depressive behavior during chronic stress.

In summary, our data show that elevated plasma levels of corticosterone suppress BDNF-TrkB protection pathway in the hippocampus, which is associated with a corresponding change in nanoscale structural disorder in the hippocampus.

**Acknowledgements**

This work was supported by NIH grants (R01EB003682 and R01EB016983, FedEx Institute, and the University of Memphis to PP